# Emergence of a noncollinear magnetic state in twisted bilayer CrI$_3$

Yang Xu[1,2], Ariana Ray[1], Yu-Tsun Shao[1], Shengwei Jiang[1], Daniel Weber[3], Joshua E. Goldberger[3], Kenji Watanabe[4], Takashi Taniguchi[4], David A. Muller[1,5], Kin Fai Mak[1,5,6*], Jie Shan[1,5,6*]

[1]School of Applied and Engineering Physics, Cornell University, Ithaca, NY, USA
[2]Beijing National Laboratory for Condensed Matter Physics, Institute of Physics, Chinese Academy of Sciences, Beijing, China
[3]Department of Chemistry and Biochemistry, The Ohio State University, Columbus, OH, USA
[4]National Institute for Materials Science, Tsukuba, Japan
[5]Kavli Institute at Cornell for Nanoscale Science, Ithaca, NY, USA
[6]Laboratory of Atomic and Solid State Physics, Cornell University, Ithaca, NY, USA

*Email: kinfai.mak@cornell.edu; jie.shan@cornell.edu

**The emergence of two-dimensional (2D) magnetic crystals and moiré engineering has opened the door for devising new magnetic ground states via competing interactions in moiré superlattices [1-12]. Although a suite of interesting phenomena, including multi-flavor magnetic states [7], noncollinear magnetic states [4,7], moiré magnon bands and magnon networks [8], has been predicted, nontrivial magnetic ground states in twisted bilayer magnetic crystals have yet to be realized. Here, by utilizing the stacking-dependent interlayer exchange interactions in CrI$_3$ (Ref. [13,14]), we demonstrate in small-twist-angle bilayer CrI$_3$ a noncollinear magnetic ground state. It consists of both antiferromagnetic (AF) and ferromagnetic (FM) domains and is a result of the competing interlayer AF coupling in the monoclinic stacking regions of the moiré superlattice and the energy cost for forming AF-FM domain walls. Above the critical twist angle of ~ 3°, the noncollinear state transitions abruptly to a collinear FM ground state. We further show that the noncollinear magnetic state can be controlled by gating through the doping-dependent interlayer AF interaction. Our results demonstrate the possibility of engineering moiré magnetism in twisted bilayer magnetic crystals, as well as gate-voltage-controllable high-density magnetic memory storage.**

Moiré superlattices built on twisted bilayers of van der Waals' materials have presented an exciting platform for studying correlated states of matter with unprecedented controllability [15,16]. In addition to graphene and transition metal dichalcogenide moiré materials [9–12,17–19], recent theoretical studies have predicted the emergence of new magnetic ground states in twisted bilayers of 2D magnetic crystals [4,7,8]. These states are originated from the stacking-dependent interlayer exchange interactions in magnetic moiré superlattices. Two-dimensional CrI$_3$, in which stacking-dependent interlayer magnetic ground states have been demonstrated by recent experiments [13,14], is an excellent candidate for exploring moiré magnetism.

There is a one-to-one correspondence between the stacking structure and the magnetic ground state in bilayer $CrI_3$: the monoclinic (M) phase supports an A-type AF ground state, in which two FM $CrI_3$ monolayers are antiferromagnetically coupled; and the rhombohedral (R) phase supports a FM ground state [3,13,14,20]. The interlayer FM exchange in the R-phase is substantially stronger than the AF exchange in the M-phase [3]. In twisted bilayer $CrI_3$, the triangular moiré superlattice has both M- and R-stacking regions (Fig. 1a), and the competing interlayer AF and FM interactions can induce nontrivial magnetic ground states [7,8]. In particular, noncollinear magnetic ground states consisting of periodic AF and FM domains (Fig. 1b) are expected, provided that the energy gain from forming the AF domains in the M-stacking regions exceeds the energy cost from forming the AF-FM domain walls. Since the energy gain scales with the area of the moiré unit cell whereas the energy cost scales with the moiré period (see below), a noncollinear magnetic ground state is favored below the critical twist angle (or above the critical moiré period) [7].

We fabricate twisted bilayer $CrI_3$ by the "tear-and-stack" method that has been widely used in making twisted bilayer graphene samples [21–23]. A series of samples with varying twist angle $\theta$ has been studied. All of the samples are encapsulated between hexagonal boron nitride (hBN) substrates for protection from environmental degradation. We have also fabricated dual-gated field-effect devices out of a subset of these samples. These devices allow us to continuously tune the interlayer AF exchange interaction in the M-stacking regions by varying the doping level in $CrI_3$ (Ref. [24,25]).

We characterize the structure of one of the samples with a target twist angle, $\theta = 1.5^0$, by scanning transmission electron microscopy (STEM). For this purpose, we place the twisted sample (encapsulated by two hBN flakes with thickness less than 5 nm) on a TEM window made of an 8-nm-thick amorphous $SiO_2$ membrane. Figure 1c shows the diffraction pattern; diffraction spots from both hBN and $CrI_3$ are observed. The third-order diffraction spots from twisted $CrI_3$ (circled) correspond to a Fourier-transformed distance of $a/3 = 0.23$ nm ($a$ is the in-plane lattice constant of $CrI_3$). Because of the finite twist angle, each $CrI_3$ diffraction spot consists of a pair of shifted diffraction spots, producing a slightly elongated spot. Figure 1d-1f show the dark-field images each constructed from a single circled diffraction spot in Fig. 1c. A clear moiré stripe pattern with periodicity one third of the moiré period $a_M/3 = 6.1 \pm 0.9$ nm is seen. Each image is related to the other by a 60° rotation. The real-space moiré pattern of the twisted sample can be constructed by superimposing the three images (Extended Data Figure 2). Using the measured $a_M = 18.3 \pm 2.7$ nm, we determine the twist angle to be $\theta = 2\sin^{-1}\left(\frac{a}{2a_M}\right) \approx 2.2 \pm 0.3^0$. The result is consistent with the typical twist angle accuracy ($\pm 0.5^0$) in these fabrication processes.

We probe the magnetic ground state of twisted bilayer samples of $CrI_3$ by magnetic circular dichroism (MCD) measurements. Under our experimental conditions the MCD is linearly proportional to the out-of-plane component of the magnetization. However, a direct comparison of the absolute MCD for different samples is not appropriate because of the different local field factors from different substrate thicknesses. Unless otherwise specified, all measurements are performed at 4 K. (See Methods for details on device fabrication and characterizations.)

Figure 2a-2d show the MCD as a function of out-of-plane magnetic field $B$ for four bilayer samples. (See Extended Data Fig. 3 for additional samples.) A natural bilayer CrI$_3$ sample (Fig. 2a) is included as a reference. In the natural bilayer the MCD is negligible till the field reaches the critical value $B_c \approx 0.6$ T for the spin-flip transition; the behavior is fully consistent with the A-type AF state reported in the literature [2,26]. The magnetic response of twisted bilayers is diametrically different. Both AF-like and FM-like behaviors are present in the 1.2°-twist sample (Fig. 2b). In addition to the spin-flip transition near $B_c \approx 0.5 - 0.6$ T, a FM loop centered at zero field emerges. (The MCD from a nearby isolated monolayer region, shown in red, is included as a comparison; it shows a clear FM loop as expected.) The result clearly demonstrates the coexistence of AF and FM domains (within a diffraction-limited laser spot of ~ 500 nm in diameter) in the 1.2°-twist sample.

With increasing twist angle $\theta$, the AF contribution quickly diminishes: a very small AF contribution remains in the 4°-twist sample; only the FM contribution is observed in the 15°-twist sample. The fraction of the FM and AF regions in the sample can be approximated by the ratio of the remnant magnetization to the saturation magnetization, $f_{FM} \approx \frac{MCD(0\ T)}{MCD(1\ T)}$, and $f_{AF} \approx 1 - f_{FM}$, respectively. Here we have ignored the domain walls since they are estimated to be narrow compared to the moiré period at low temperatures in twisted bilayer CrI$_3$ (Methods). The twist angle dependence of $f_{FM}$ is summarized in Fig. 3 (top panel). The AF-FM coexistence disappears at a critical angle $\theta_C \sim 3°$, above which the entire sample is FM. In addition, below the critical angle the spin-flip transition field $B_c$ is weakly angle dependent (lower panel).

We characterize the spatial homogeneity of the twisted bilayer samples by MCD imaging. Figure 2e and 2f show $[MCD(1\ T) - MCD(0\ T)]$ and $MCD(0\ T)$, respectively, to represent the AF and FM contributions in the 1.2°-twist sample (the boundary of each CrI$_3$ monolayer is marked by dashed curves). We observe $f_{AF} \approx 0$ in the isolated monolayer regions as expected, and a finite $f_{AF}$ in most of the twisted bilayer regions, demonstrating the robustness of the AF-FM coexistence in small-twist-angle bilayers. The FM contribution in the bilayer regions is comparable to that of the isolated monolayers. This is also consistent with the AF-FM coexistence that reduces the overall FM contribution to the $MCD(0\ T)$ signal. On the other hand, in twisted bilayers with $\theta > \theta_c$ the AF contribution is absent over the entire sample (Extended Data Fig. 4), and the FM signal nearly doubles that of the isolated monolayer regions (Fig. 2g for $\theta = 4°$, Fig. 2h for $\theta = 15°$).

We further examine the temperature dependence of the magnetic response of the 1.2°-twist sample in Fig. 4. Figure 4a shows the magnetic-field dependence of the MCD (scanned from -1.5 T to 1.5 T) at varying temperatures $T$ in a contour plot. The FM and AF fractions ($f_{FM}$ and $f_{AF}$) are nearly temperature independent in the examined temperature range. Field dependences at representative temperatures are shown in Fig. 4b for both field scan directions. Upon cooling, the coercive field of the FM contribution increases monotonically because the magnetic anisotropy increases. Intriguingly, the spin-flip transition field $B_c$ first increases, followed by a small reduction when the temperature further decreases. This is in stark contrast to the monotonic increase of $B_c$ with decreasing temperature in natural bilayer $CrI_3$ (Extended Data Fig. 5). The temperature dependences of the FM coercive field and $B_c$ are summarized in Fig. 4c.

The observed AF-FM coexistence in small-twist-angle $CrI_3$ bilayers is consistent with the emergence of a noncollinear magnetic ground state driven by competing magnetic interactions in a moiré superlattice [7]. The moiré superlattice of twisted bilayer $CrI_3$ contains spatially modulated M- and R-stacking structures (Fig. 1a), which support the AF and FM interlayer exchange coupling, respectively. Because the interlayer FM exchange in the R-phase is substantially stronger than the interlayer AF exchange in the M-phase [3], we only need to consider the competition between the formation of the AF domains and the AF-FM domain walls. The energy gain per moiré unit cell from forming the AF domains in the M-stacking regions, $\sim 2 f_{AF} \left(\frac{a_M}{a}\right)^2 J_\perp$, scales with the area of the moiré unit cell. The energy cost per moiré unit cell from forming the AF-FM domain walls near the M-R stacking boundaries, $\sim \pi \left(\frac{a_M}{a}\right) \sqrt{J_\parallel (K + 2J_\perp)}$, scales with the period of the moiré unit cell (see Methods for derivations). Here $J_\perp \sim 0.1$ meV is the interlayer AF exchange constant [2,26], $J_\parallel \sim 2$ meV is the intralayer FM exchange constant [27], and $K \sim 0.3$ meV is the single-ion magnetic anisotropy energy in $CrI_3$ (Ref. [28]). The different scaling of the two energy terms with moiré period guarantees that at small twist angle (or large $a_M$) the energy gain from forming the AF domains wins and a noncollinear magnetic ground state with spatially modulated AF and FM regions emerges (Fig. 1b). The critical twist angle can be estimated by equating the two energy terms to yield $\theta_c \sim \frac{f_{AF} J_\perp}{\sqrt{J_\parallel (K + 2J_\perp)}} \approx 3°$. The value is in good agreement with experiment. Above the critical angle, a collinear FM ground state is preferred.

The transition from a noncollinear to a collinear magnetic state tuned by the twist angle can be continuous or abrupt depending on the magnetic anisotropy. It is predicted to be continuous for small $K/J_\perp$ and abrupt for large $K/J_\perp$ (Ref. [7]). In the continuous phase transition scenario, an intermediate noncollinear state with spins flop to the in-plane direction in the AF or FM domains is further expected. In $CrI_3$ with large magnetic anisotropy ($K/J_\perp \sim 3$), the intermediate state has a high energy cost and an abrupt transition is favored. The absence of an intermediate state is consistent with the presence of spin-flip transitions in all small-twist-angle bilayers and with the observed weak angle dependence of $B_c$ for $\theta < \theta_c$ (Fig. 3). Because of the large magnetic anisotropy, the domain wall width is small compared to $a_M$ so that $B_c$ is weakly angle dependent for $\theta < \theta_c$ (Methods).

The non-monotonic temperature dependence of $B_c$ is also consistent with the coexistence of the AF and FM domains. In natural bilayer $CrI_3$, the spin-flip transition occurs at $B_c \sim \frac{2 J_\perp}{3 \mu_B}$ (Ref. [26]), at which the Zeeman energy gain $3\mu_B B$ of the FM state overcomes the interlayer AF exchange energy cost $2 J_\perp$. Here $\mu_B$ is the Bohr magneton and each Cr ion carries $3\mu_B$. In twisted bilayer CrI$_3$, the presence of AF and FM domains lowers the energy cost for the spin flip transition by the domain wall energy and gives rise to a small reduction in the spin-flip transition field $B_C \sim \frac{2f_{AF}J_\perp - \pi\left(\frac{a}{a_M}\right)\sqrt{J_\parallel(K+2J_\perp)}}{f_{AF}(3\mu_B)}$ (see Methods for derivation). The observed non-monotonic temperature dependence of $B_C$ can be explained by the different scaling of the exchange constants $J_\parallel, J_\perp \propto t^\alpha$ (with $\alpha < 1$) and the anisotropy energy $K \propto t^\beta$ (with $\beta > 1$) with the reduced temperature $t = \frac{T_c - T}{T_c}$ ($T_c$ is the critical temperature) [29–31]. Using $\alpha = 0.22$ (Ref. [27]) for both $J_\parallel$ and $J_\perp$ and $\beta = 2.3$, we can reproduce the observed non-monotonic temperature dependence of $B_c$ in Fig. 4c (red line).

Finally, we demonstrate gate control of the noncollinear magnetic state. Figure 5a shows the magnetic-field dependence of MCD (normalized by the saturated value at 1 T) at varying gate voltages for the 1.2°-twist sample. With increasing gate voltage (corresponding to electron doping), the spin-flip transition field $B_c$ decreases by as much as ~ 50% while the FM loop is affected less. The result shows that the primary effect of electron doping is to weaken the interlayer AF exchange $J_\perp$ in the M-stacking regions. This is consistent with earlier reports on natural bilayer CrI$_3$ (Ref. [24,25]). Meanwhile, the remnant magnetization as manifested by $MCD(0\,T) \propto f_{FM}$ increases with gating. The gate voltage dependence of $f_{FM}$ is shown in Fig. 5b. It increases by ~ 10% with electron doping. (The reproducibility on other field effect devices is shown in Extended Data Fig. 6.)

The increase in the remnant magnetization with electron doping reflects shrinkage of the AF domains (or expansion of the FM domains) as the interlayer AF exchange constant $J_\perp$ is weakened. At sufficiently high electron doping densities, where $J_\perp$ becomes smaller than the critical value $J_{\perp c} \sim \theta\sqrt{J_\parallel K}/f_{AF}$ (Methods), the energy gain from forming AF domains at the M-stacking regions would become smaller than the energy cost of forming AF-FM domain walls, and an abrupt quantum phase transition from a noncollinear state to a collinear FM state would emerge [7]. Tuning $J_\perp$ by electron doping in a sample with fixed twist angle $\theta$ is therefore similar to varying $\theta$ while keeping $J_\perp$ constant (Fig. 3). Future studies with higher electron doping densities will provide a promising route to achieve a gate-voltage-induced magnetic quantum phase transition.

In conclusion, we have realized moiré superlattices of 2D magnetic crystals and demonstrated the emergence of a noncollinear magnetic ground state in small-twist-angle bilayer CrI$_3$. The noncollinear state is a result of the competing magnetic interactions in a moiré superlattice, and can be continuously tuned by electrical gating. The observations are largely consistent with the predictions of a spin model for twisted bilayer magnetic materials [7]. Our results form the basis for engineering moiré magnetism in twisted bilayer magnetic crystals and for studying gate-voltage-controlled magnetic quantum phase transitions.

## Methods
### Device fabrication
We fabricated twisted bilayer $CrI_3$ devices by the "tear-and-stack" method [21–23] inside a glove box with water and oxygen levels below 1 part per million (ppm). Monolayer $CrI_3$ was mechanically exfoliated from bulk $CrI_3$ crystals onto $SiO_2$/Si substrates and identified by its optical contrast under a microscope. One example is shown in Extended Data Fig. 1a. We picked up a part of the flake (the black dashed curve) from the substrate with a polydimethylsiloxane/polycarbonate (PDMS/PC) stamp and then twisted the remaining part (the red dashed curve in Extended Data Fig. 1b) on the substrate by an angle $\theta$ that ranges from 0° to 60°. A fine angle control is achieved by using a high-precision rotation stage (Thorlabs PR01). Finally, we engaged the two monolayers to form a twisted bilayer (Extended Data Fig. 1c). The finished twisted bilayer was encapsulated by hBN and contacted with graphene electrodes. A subset of the samples also has extra few-layer graphene as top and bottom gate electrodes. For the STEM studies, we transferred the hBN-encapsulated samples onto a TEM window made of an 8-nm-thick amorphous $SiO_2$ membrane.

### MCD measurements
Similar MCD measurements on 2D $CrI_3$ have been reported in earlier studies [13,24,26]. In short, an incident He-Ne laser beam (633 nm) was focused onto the sample plane to a diffraction-limited spot (~ 500 nm in diameter) using a microscope objective with high numerical aperture (N.A. ~ 0.81). The polarization of the incident light was modulated between left and right circular polarization by a photoelastic modulator. The reflected light was collected by the same objective and directed to a biased photodiode. The AC and DC components of the reflected light were collected by a lock-in amplifier and a multi-meter, respectively. The MCD signal is defined as the ratio of the AC to the DC signal. For MCD imaging studies, we used a filtered tungsten halogen lamp output (centered at 633 nm with a bandwidth ~ 10 nm) for broad-field illumination in the same setup. The reflected light intensity from left- and right-handed illumination was collected to a charge-coupled device (CCD) that produces an optical image of the sample. The MCD image at a given magnetic field is obtained by normalizing the difference between the two images by their sum.

### 4D-STEM imaging and image analysis
The twisted bilayer $CrI_3$ sample was imaged in an FEI F20 S/TEM at 200 kV using a 10 μm condenser aperture and a probe semi-convergence angle of ~1.5 mrad. The sample was imaged using the 4D-STEM technique, in which an electron probe is stepped over a scan region and a full electron diffraction pattern is recorded at each pixel step. The final 4D dataset (2D diffraction pattern x 2D image) was then used to recover details about the local crystal structure over large fields of view. All 4D-STEM imaging was done using the Electron Microscope Pixel Array Detector (EMPAD) [32]. The high dynamic range of the EMPAD ($10^6$:1) allows for the collection of both the direct and scattered electron beams. A real-space scan size of 256x256 pixels and an acquisition time of 3 ms/pixel were used.

The 4D-STEM data were screened and processed using the Cornell Spectrum Imager [33]. Each 4D-STEM dataset consists of a 2D real-space scan and a 2D diffraction pattern associated with each real-space pixel in that scan. Virtual dark field images are created by masking a region of the diffraction pattern and summing over the contributions of each pixel in the mask to obtain the intensity of the real-space image. The average moiré periodicity was calculated from the fast Fourier transform (FFT) of one of the dark field images. The FFT contains a pair of peaks arising from the real-space moiré. A line profile was taken through the FFT and Gaussian functions were fitted to the peaks in MATLAB. The moiré periodicity and uncertainty were taken to be the mean and one standard deviation from the Gaussian fit parameters. Using this method, the moiré period in Figure 1d-1f was found to be 18.3 ± 2.7 nm.

**AF-FM domain wall width**

We consider a model that involves an abrupt physical boundary between the monoclinic and rhombohedral stacking regions in the moiré superlattice of twisted bilayer $CrI_3$. The simple model is intended to provide an order of magnitude estimate of the experimental observations. Since the interlayer FM exchange interaction for rhombohedral stacking is predicted to be substantially stronger than the interlayer AF exchange interaction for monoclinic stacking [3], we assume the AF-FM domain walls can only form in the monoclinic stacking regions. The domain wall energy per moiré unit cell consists of three terms:

$$E_{DW} \sim \pi^2 \left(\frac{a_M}{\delta}\right)\frac{J_\parallel}{2} + \left(\frac{\delta a_M}{2a^2}\right)(K + 2J_\perp). \tag{1}$$

Here $\delta$ is the domain wall width and $\frac{\delta a_M}{a^2} \sim N$ is the number of spins in the domain wall (per moiré unit cell). The first two terms, involving intralayer FM exchange $J_\parallel$ and magnetic anisotropy $K$, constitute the standard expression for the energy cost of forming domain walls in FM materials. They compete against each other; a larger $J_\parallel$ favors a wider domain wall; conversely, a larger $K$ favors a narrower domain wall. The last term is the energy cost from interlayer exchange interactions in the monoclinic stacking regions. It can be evaluated from $\sum_{i=1}^{N} J_\perp (1 + \cos\phi_i) \approx \frac{NJ_\perp}{\pi} \int_0^\pi d\phi\,(1 + \cos\phi)$, where $\phi$ is the spin angle between the top and bottom layer. Minimizing $E_{DW}$ with respect to $\delta$ then yields an expression for the domain wall width

$$\delta \sim \pi \sqrt{\frac{J_\parallel a^2}{K + 2J_\perp}}. \tag{2}$$

It is analogous to the result for domain walls in FM materials, $\delta \sim \pi \sqrt{\frac{J_\parallel a^2}{K}}$, with the modification of $K \to K + 2J_\perp$. We estimate the domain wall width in $CrI_3$ to be about 4 nm at low temperatures, which is substantially smaller than $a_M \sim 20$ nm. At higher temperatures (still below $T_C$), where the magnetic anisotropy energy $K$ becomes negligible compared to $J_\perp$, the domain wall width becomes $\delta \sim \pi \sqrt{\frac{J_\parallel a^2}{2J_\perp}}$. Therefore, $\delta$ here is expected to have a much weaker temperature dependence compared to FM materials.

**Critical twist angle and interlayer exchange coupling**
In the above model, the noncollinear magnetic ground state is originated from the competing magnetic interactions between the formation of AF-FM domain walls and interlayer AF exchange in the monoclinic stacking regions. A noncollinear state is favored only if the latter dominates. The energy gain per moiré unit cell from having AF domains in the monoclinic stacking regions is:

$$E_{AF} \sim 2 f_{AF} \left(\frac{a_M}{a}\right)^2 J_\perp. \tag{3}$$

The energy cost per moiré unit cell from forming the domain walls, after substituting (2) into (1), is:

$$E_{DW} \sim \pi \left(\frac{a_M}{a}\right) \sqrt{J_\parallel (K + 2J_\perp)}. \tag{4}$$

The noncollinear-to-collinear phase transition occurs when $E_{AF} \sim E_{DW}$. If the transition is tuned by the twist angle $\theta$, we can obtain a critical angle, by using $\frac{a}{a_M} = 2 \sin\frac{\theta}{2} \approx \theta$,

$$\theta_c \sim \frac{f_{AF} J_\perp}{\sqrt{J_\parallel(K+2J_\perp)}}. \tag{5}$$

Alternatively, if the transition is tuned by varying $J_\perp$ through gating at a fixed twist angle $\theta$, we can obtain a critical interlayer exchange, by keeping the leading order term,

$$J_{\perp c} \sim \theta \sqrt{J_\parallel K}/f_{AF}. \tag{6}$$

**Spin-flip transition field**
The spin-flip transition field $B_c$ for the AF domains in the noncollinear phase can be evaluated following similar consideration of the energetics. The application of an external magnetic field favors the collinear state. The magnetic energy per moiré unit cell of the monoclinic stacking regions for the collinear state is

$$E_{col} \sim f_{AF} \left(\frac{a_M}{a}\right)^2 (J_\perp - 3\mu_B B). \tag{7}$$

The first term is the energy cost for having an interlayer FM configuration in the collinear state; and the second term is the Zeeman energy gain for the FM state. On the other hand, the energy per moiré unit cell at the monoclinic stacking regions for the noncollinear state is

$$E_{ncol} \sim E_{DW} - f_{AF} \left(\frac{a_M}{a}\right)^2 J_\perp. \tag{8}$$

The first term is the energy cost for forming domain walls and the second term is the energy gain for having an interlayer AF configuration in the monoclinic stacking regions. The spin-flip transition field can be obtained by equating (7) and (8):

$$B_c \sim \frac{2f_{AF}J_\perp - \pi\left(\frac{a}{a_M}\right)\sqrt{J_\parallel(K+2J_\perp)}}{f_{AF}(3\mu_B)}. \tag{9}$$

**Figures and figure captions**

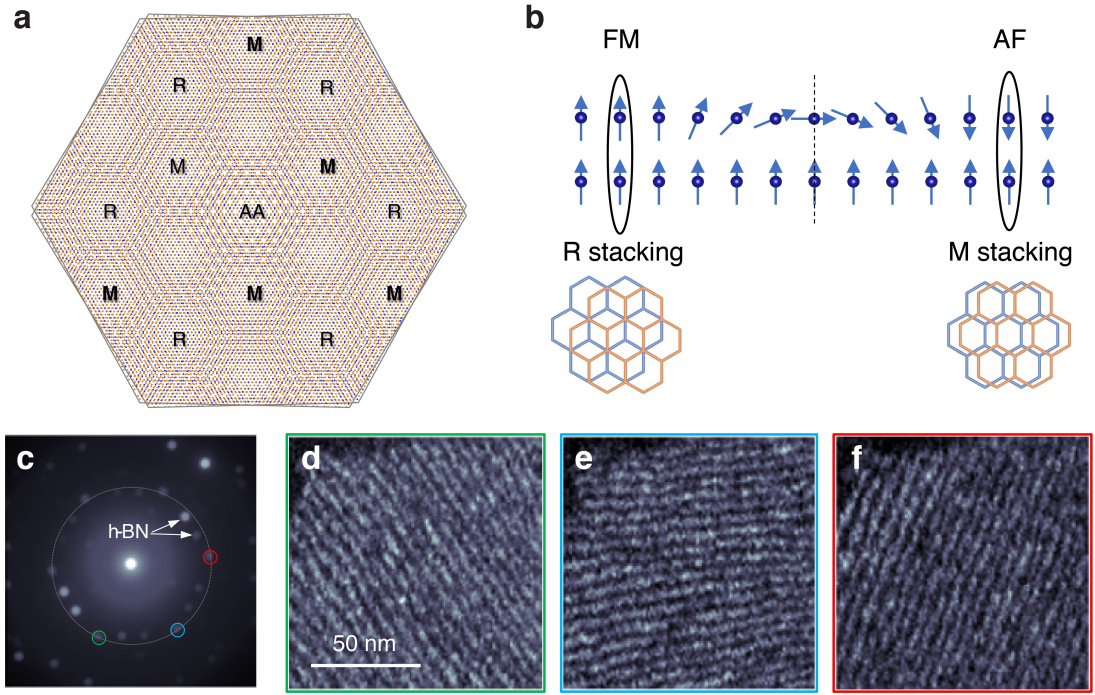

**Figure 1 | Moiré superlattice structure of twisted bilayer CrI$_3$. a**, Moiré superlattice structure of twisted bilayer CrI$_3$ with a small twist angle $\theta$. R, M and AA denote rhombohedral, monoclinic and AA stacking, respectively. **b**, Schematic illustration of a magnetic domain wall formed between the R and M stacking regions. **c**, Electron-beam diffraction pattern from a small-angle twisted bilayer CrI$_3$ sample, down the <001> zone axis (perpendicular to sample surface). The two arrows indicate the diffraction spots (six pairs in total) from the top and bottom hBN capping layers. The big dashed circle tracks the <300> diffraction spots for CrI$_3$. The spots circled in green, blue and red are the dark field masks for the real-space dark field moiré fringe patterns in **d-f**, respectively.

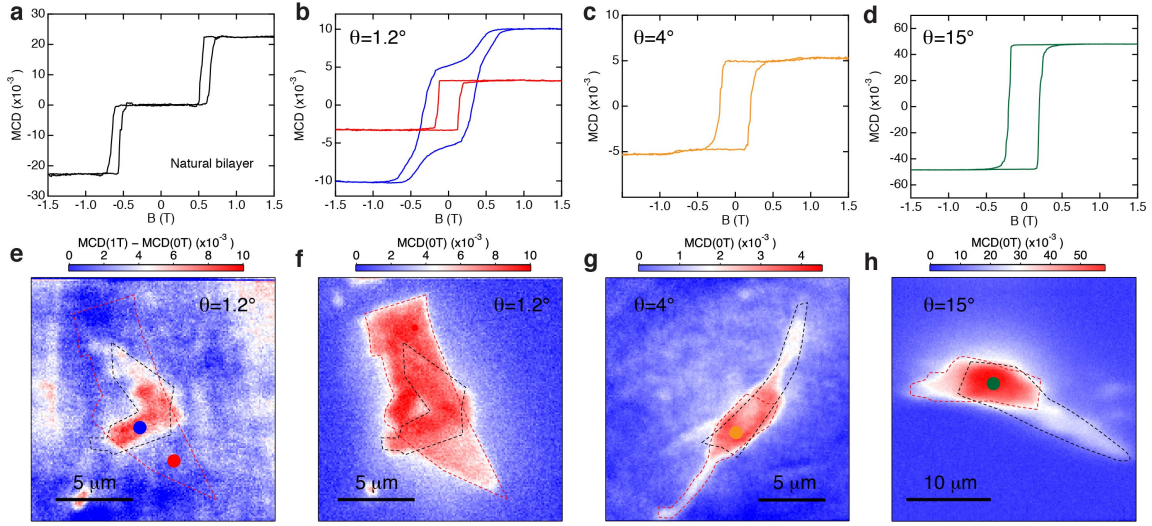

**Figure 2 | MCD microscopy in twisted bilayer CrI$_3$**. **a-d**, Magnetic field dependence of MCD for natural bilayer CrI$_3$ (**a**) and twisted bilayer CrI$_3$ with twist angle 1.2° (**b**), 4° (**c**) and 15° (**d**). The MCD signal of an isolated monolayer CrI$_3$ is also shown in **b** for comparison. Coexistence of AF and FM states is seen at small twist angles. **e**, Image of $MCD(1\,T) - MCD(0\,T)$ for the 1.2° sample. It shows the AF fraction of the sample. Non-zero contrast is seen only at the twisted bilayer region. **f-h**, MCD images at $B = 0$ T for the 1.2° (**f**), 4° (**g**) and 15° (**h**) samples after field polarization at 1 T. They show the FM fraction of the samples. In all images the dashed black and red lines outline the constituent monolayer regions. The color spots denote the locations of the MCD measurements in **b-d** with the same color.
12

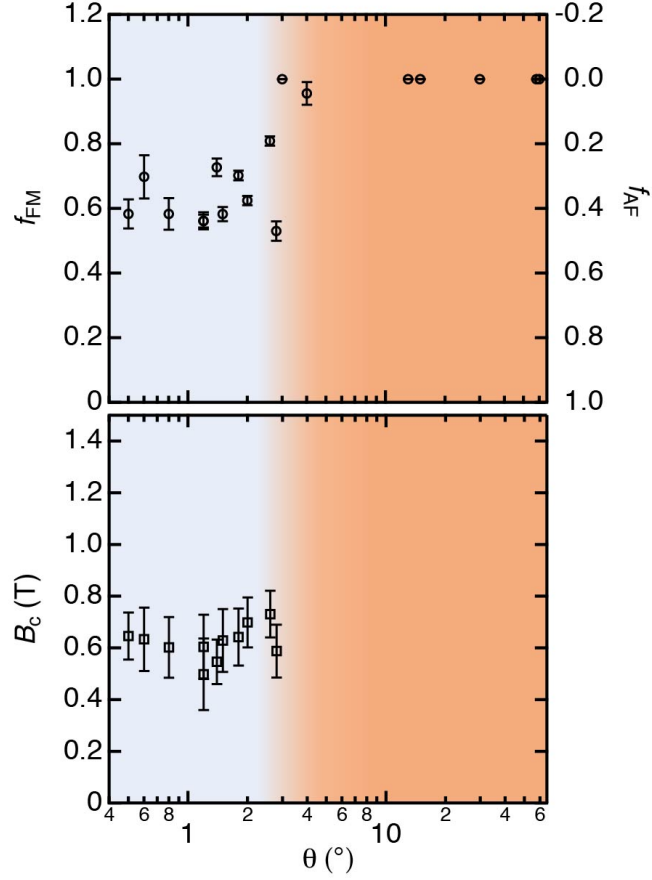

**Figure 3 | Dependence on twist angle**. Upper panel: summary of $f_{FM}$ (left axis) and $f_{AF}$ (right axis) as a function of twist angle $\theta$ for all the measured samples. The error bars are estimated from measurements at different spots in the same sample and indicate the spatial inhomogeneity. The twist angle error bar (not shown) is estimated to be about 0.5°. $f_{FM} <1$ at small twist angles ($\theta \lesssim 3°$) indicates a mixed AF-FM ground state whereas a pure FM ground state is observed at large twist angles ($\theta \gtrsim 4°$). Lower panel: the dependence of the spin-flip transition field $B_c$ on $\theta$. The $B_c$ error bars are estimated from the field span of the spin-flip transition [26].

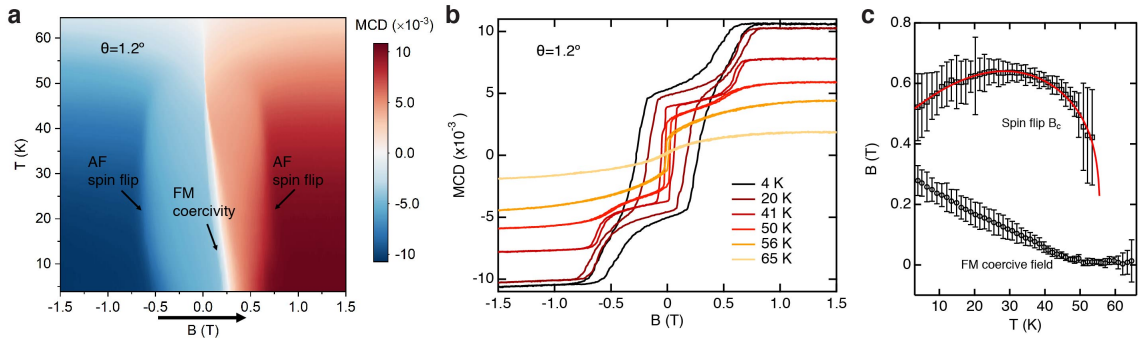

**Figure 4 | Temperature dependence. a**, MCD as a function of magnetic field (sweeping from negative to positive) and temperature for a 1.2°-twist bilayer $CrI_3$. **b**, Magnetic-field dependence of the MCD (for both forward and backward sweeps) at representative temperatures. **c**, The extracted spin-flip transition field $B_c$ and FM coercive field as functions of temperature. $B_c$ is the average spin-flip transition field between forward and backward sweeps. The red curve is a fit to $B_c$ using the model described in the main text. The vertical error bars are estimated from the field span of the magnetic transitions.

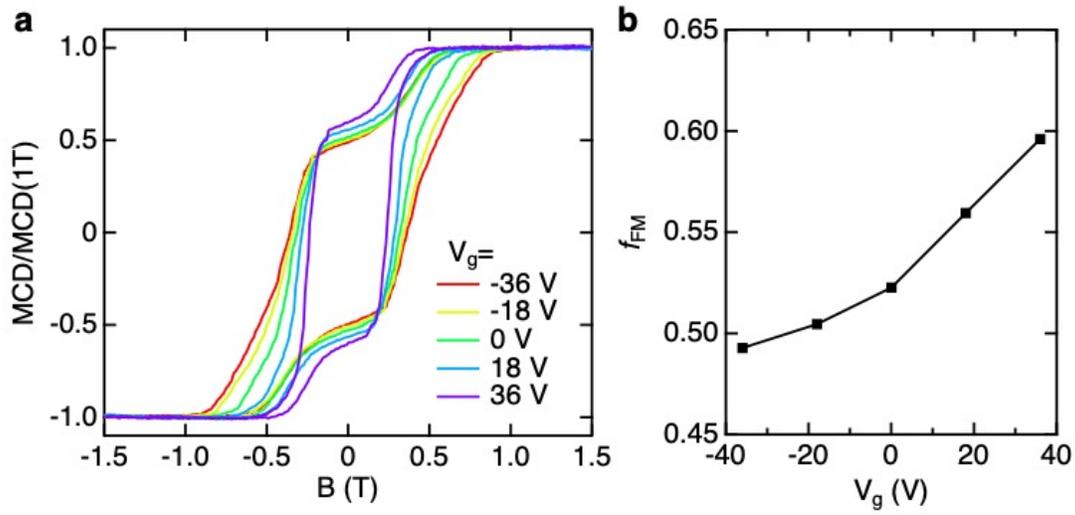

**Figure 5 | Gate control of the noncollinear magnetic state. a**, Normalized MCD as a function of magnetic field at selected gate voltages ($V_g$). $V_g$ is the total gate voltage applied symmetrically on the top and bottom gates. **b**, The extracted FM fraction $f_{FM}$ as a function of $V_g$.

**Extended Data Figures**

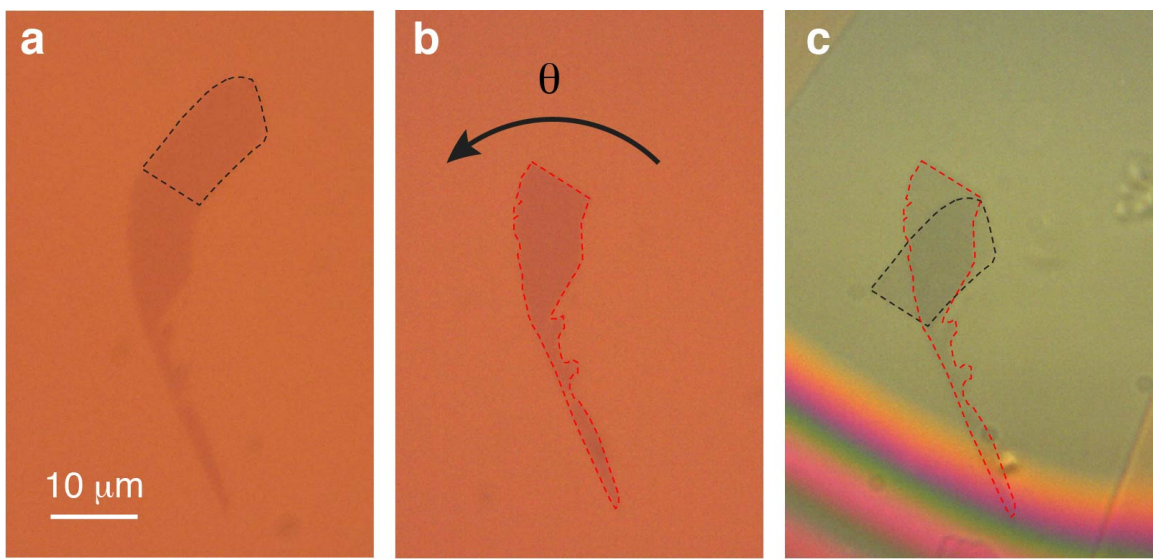

**Extended Data Figure 1 | "Tear-and-stack" optical images. a-c**, Optical images taken during the preparation of a twisted bilayer CrI$_3$ sample. The monolayer in **a** is torn apart partially (outlined by the black dashed curves) and picked up by an hBN flake, leaving the remaining portion on the substrate. The remaining portion in **b** (outlined by the red dashed curves) is rotated by an angle $\theta$. The two monolayers of CrI$_3$ are finally engaged to form a twisted bilayer (**c**).

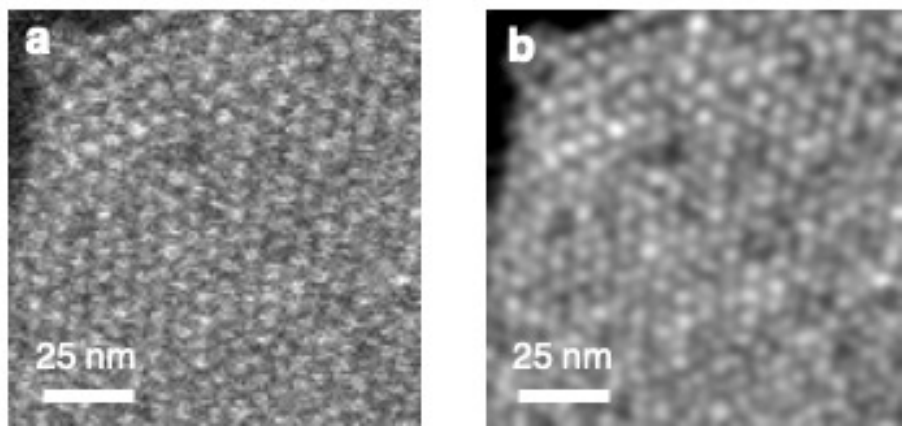

**Extended Data Figure 2 | Sum of the STEM dark field images in Fig. 1d-1f before (a) and after (b) applying a sigma=1.25 pixel Gaussian filter.** Since the image is constructed from the third-order diffraction spots from twisted CrI$_3$, the true moiré period is three times of the value shown by this figure.

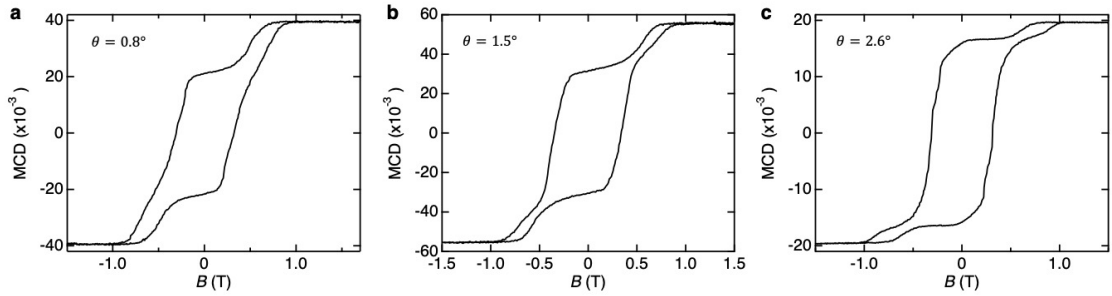

**Extended Data Figure 3 | Additional MCD data at 4 K for samples with twist angles 0.8° (a), 1.5° (b) and 2.6° (c).**

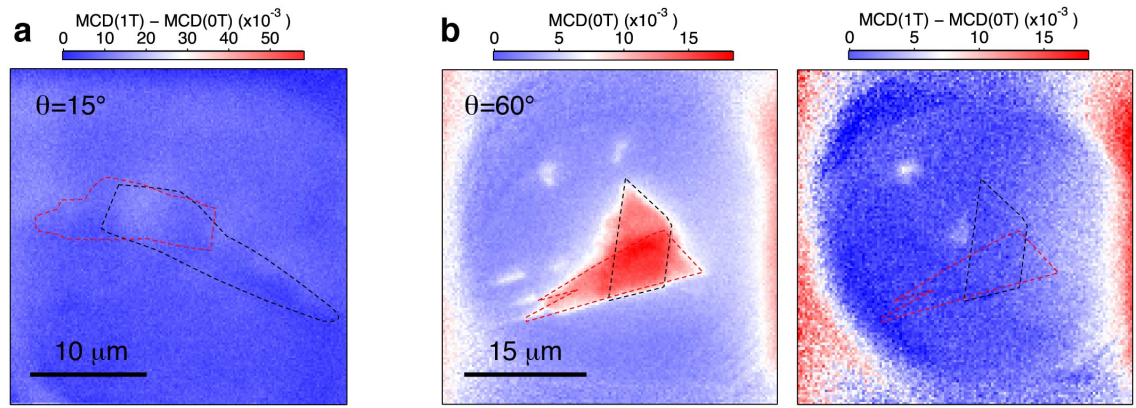

**Extended Data Figure 4 | Additional MCD imaging data**. **a**, $MCD(1\,T) - MCD(0\,T)$ map for 15° twisted bilayer CrI$_3$. No AF contribution is observed as expected. **b**, $MCD(0\,T)$ (left) and $MCD(1\,T) - MCD(0\,T)$ (right) for a 60° sample. Only FM contribution is seen. The black and red dashed lines outline the constituent CrI$_3$ monolayers. The $MCD(0\,T)$ image was taken after the sample is polarized at B = 1 T.

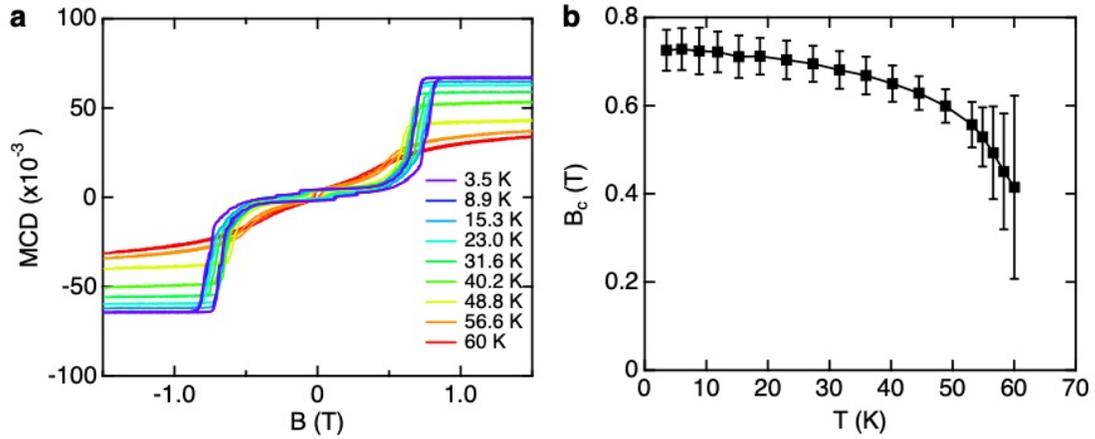

**Extended Data Figure 5 | Temperature dependence in natural bilayer CrI$_3$.** MCD response of a natural bilayer CrI$_3$ at different temperatures (**a**) and the temperature dependence of the spin flip transition field $B_c$ (**b**). $B_c$ is the average spin-flip transition field between forward and backward sweeps. The error bars are estimated from the width of the transition.

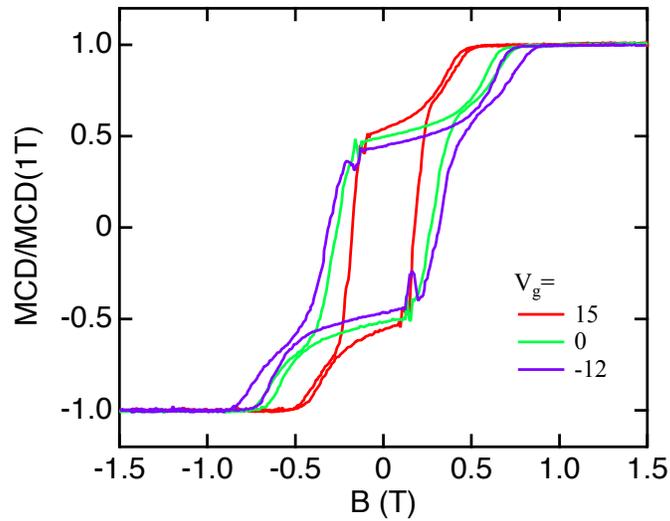

**Extended Data Figure 6 | Additional gate-dependent MCD response at 4 K for another sample with $\theta = 1.6°$.**

18